\documentclass{elsart}

\begin{document}

\begin{frontmatter}

\title{Mean-Field Study of the Degenerate Blume-Emery-Griffiths
Model in a Random Crystal Field\thanksref{cnpq}}

\author[UFSC]{N. S. Branco\thanksref{main}} and
\author[CNEN]{Luciano Bachmann}

\address[UFSC]
{Universidade Federal de Santa Catarina, Depto. de F\'{\i}sica \\
88040-900, Florian\'opolis, SC, Brazil, e-mail: nsbranco@fisica.ufsc.br}
\address[CNEN]{CNEN - IPEN - Caixa Postal 11049, 05422-970, S\~ao Paulo,
SP, Brazil} 
\thanks[cnpq]{Work partially supported by the Brazilian Agencies 
CNPq and FINEP.}
\thanks[main]{Author to whom the proofs should be sent.}

\begin{abstract}
The degenerate Blume-Emery-Griffiths (DBEG) model has recently been introduced
in the study of martensitic transformation problems. 
This model has the same Hamiltonian as the standard Blume-Emery-Griffiths 
(BEG) model but, to take into
account vibrational effects on the martensitic transition, it is
assumed that the states $S=0$ have a degeneracy $p$ ($p=1$ corresponds to
the usual BEG model). 
In some materials, the transition would be better described by a 
disordered DBEG model; further, the inclusion of disorder in the DBEG model 
may be relevant in the study of shape memory alloys.
From the theoretical point of view, it would be interesting
to study the consequence of conflicting effects: the parameter $p$, which
increases the first-order phase-transition region, and disorder in the 
crystal field, which tends to diminish this region in three dimensions.
In order to study this competition in high-dimensional systems, we apply
a mean-field approximation: it is then possible to determine the critical
behavior of the random DBEG model for any value of the interaction
parameters. Finally, we comment on (preliminary) results obtained for
a two-dimensional system, where the randomness in the crystal-field has a 
more drastic effect, when compared to the three-dimensional model.

\end{abstract}


\end{frontmatter}

\baselineskip=21.5pt

\section{Introduction}

	The hamiltonian of the degenerate Blume-Emery-Griffiths (DBEG) model 
reads: 
$ \mathcal{H} = - J \sum_{<i,j>} S_i S_j - K \sum_{<i,j>} S_i^2 S_j^2  
+ \Delta \sum_i S_i^2$,
where $J$ is the exchange constant,
$K$ is the biquadratic interaction, $\Delta$ is a crystal field,
the first two sums are over nearest-neighbor pairs on a lattice,
the third sum is over all sites, $S_i = \pm 1,0$, and the
$S=0$ states have a degeneracy $p$ \cite{barce}. 
This parameter mimics 
the effects of vibrational degrees of freedom on martensitic transitions.
The DBEG model was studied, for a particular value
of the crystal-field, within a mean-field approximation
and Monte Carlo simulation: the effect of $p$ is to shrink the ferromagnetic
phase and to increase the region where the transition is of first-order
\cite{barce}. 

	Nevertheless, some materials were found to be better described
by a disordered DBEG model, which may be relevant in the
description of conventional shape memory alloys as well \cite{barce}.
From the theoretical point of view, on the other hand, it has been shown that 
randomness may have drastic consequences on multicritical behavior. In
two dimensions, for instance, any infinitesimal amount of disorder
supresses non-symmetry-breaking first-order phase transitions and
replaces symmetry-breaking first-order phase transitions by continuous ones.
The effect of disorder on high-dimensional systems is different:
first-order phase transitions disappear only at a finite amount of
randomness \cite{berker}. This behavior has been observed in a large variety
of models, either with random ``field'' or with random ``interaction''
\cite{berker,branco,cardy}. 

	Thus, in the disordered DBEG model two ingredients have opposite
effects: while the degeneracy factor $p$ increases the region where the
transition is of first-order, disorder tends to supress this region. This
competition may have interesting effects on the phase diagrams, apart
from  potential experimental relevance. In order to study this competition on
high-dimensional systems, we apply a mean-field approximation to the
DBEG model in a random crystal-field, which follows the probability 
distribution $\mathcal {P}(\Delta_i) = r \; \delta(\Delta_i+\Delta) +
    (1-r) \; \delta(\Delta_i-\Delta)$.
It is worthy stressing that, if the interaction parameters, $J$ and $K$, 
were chosen to be random, instead of the crystal field, the overall 
consequences on the phase diagram would be the same \cite{berker}.

\section{Formalism and Results}

	Our approximation is based on the Gibbs inequality for the free
energy \cite{ajp}:
 $ F \leq \mbox{Tr} \rho \mathcal{H} + (1/\beta) \mbox{Tr} \rho \ln \rho$,
where $F$ is the exact free energy, $\mathcal{H}$ is the DBEG hamiltonian
and $\rho$ is an exactly solvable density matrix. We chose $\rho$ as
the most general single-site density matrix for the ferromagnetic
(positive $J$ and $K$) system. Since this procedure is usual,
we will not discuss it in detail.

	Let us first comment on some general results we obtain, for any
$K/J$. The transition at zero temperature is not affected by the parameter
$p$ and takes place at $(\Delta/zJ) = (K/J+1)(1+r)/2$; for smaller values
of $\Delta/zJ$ the magnetization equals 1, while $m=r$ for 
$\Delta/zJ > (K/J+1)(1+r)/2$. 

	For $\Delta/zJ=\infty$, the random field DBEG model,
with the field distribution given by the equation in the previous page, is 
equivalent
to the quenched site-diluted spin-1/2 Ising model. The states $S=\pm 1$
represent magnetic sites and the states $S=0$ represent non-magnetic
impurities. Therefore, only for $r \geq r_c$ (where $r_c$ 
is the lattice-dependent
value of $r$ where an infinite cluster of present sites is formed for
the first time, when $r$ is increased from zero) the critical line
between $O2$ and $D$ (see figures) should
extend to $\Delta/J = \infty$. For $r<r_c$, the ``tail'' of continuous
transition would touch the zero temperature axis at a finite value of
$\Delta/zJ$. The reason we find this ``tail'' extending to $\Delta/J = \infty$
for any $r \neq 0$ is that the present approximation is equivalent
to a model with infinite-range interactions (see Ref. \cite{branco}). 

	The assymptotic value of $kT/zJ$ for $\Delta/zJ = \infty$ equals
$r$. This result comes from the mapping cited in the last paragraph: the
degeneracy of the $S=0$ states and the biquadratic interaction $K$ play
no role in the dynamics of the $S=\pm 1$ states. If the distribution
$\mathcal{P}(\Delta_i) = r \delta(\Delta_i) + (1-r) \delta(\Delta_i-\Delta)$
is used, a different result will hold. The limit $\Delta/zJ = \infty$
is now equivalent to the \emph{spin-1} site-diluted Ising model; therefore,
$K$ and $p$ are relevant and the assymptotic value of $kT/zJ$ depends
on both parameters.
	
	Since for high values of $K/J$ the critical behavior does not depend
strongly on $p$, we well depict only
the phase diagrams for $K/J=0$, where qualitative changes occur when
$p$ is varied.

	In Figs. 1 and 2 we show the
$kT/zJ \times \Delta/zJ$ phase diagrams for $K/J=0$ and for some values
of $p$. For
$r=0.1$, the reentrant behavior observed for $p=1$ is lost and a tricritical 
point is introduced for $p=3$. This latter curve is qualitatively equivalent 
to the phase diagram for $r$ close to zero. For $r=0.3$ a reentrant behavior
is obtained for high enough $p$ (curve(c)); a critical end point (E), which
was not present for $p=1$ and $p=3$, is also observed. The first order
transition curve for $p=7$ extends to negative values of $\Delta/zJ$, until
it reaches a line of continuous transitions at a tricritical point (not
depicted in the Fig. 2).

\section{Conclusion}

	Using a mean-field approximation, we calculate the phase diagrams 
for the random crystal field DBEG model, varying the degree of disorder, $r$,
and the degeneracy factor $p$. For high values of $K/J$
the phase diagrams are almost insensitive to the value of $p$, while
for $K/J=0$ the dependence on $p$ is stronger. For the latter value
of $K/J$ and $r=0.3$ the phase diagram changes qualitatively when $p$ is 
increased enough:
while for $p=1$, for instance, the order-disorder transition is always
continuous and the first-order transition inside the ordered phase ends
in a critical point, for $p=7$ a line of first order transition separates the
ordered and disordered phases at low values of $\Delta/zJ$. A reentrant 
behavior is obtained
for high $p$. For $K=0$ and $r=0.1$, the degeneracy reintroduces a
tricritical point and eliminates the reentrant behavior.

	Some of the features found in this work may be artifact of
the simple mean-field approximation we used and should not hold
for a low-dimensional system.  Preliminary results for
the square lattice, using a real-space renormalization-group approach,
show a different behavior: for any infinitesimal amount of disorder the
first order transition is supressed, independent of $p$.

\newpage

\begin{center}
FIGURE CAPTIONS
\end{center}

	Figure 1: Phase diagram of the random-field DBEG model, for $K/J=0$
and $r=0.1$.
Dashed (continuous) lines indicate first-order (continuous) transition,
$k$ is the Boltzman constant, $T$ is the temperature and $z$ is the
coordination numer of the lattice. $O1(O2)$ stands for the ordered phase 
with $m=1(m=r)$ at $T=0$, 
$D$ stands for the disordered phase, $C$ stands for critical points,
$E$ stands for critical end points, and $TC$ stands for tricritical
points. Finally, $(a)$ stands for $p=1$, while $(b)$ stands for $p=3$.
The diagram for $r=0$ is qualitatively the same as the curve $(b)$,
except for the $O2$ phase, which is no longer present.

	Figure 2: Phase diagram of the random-field DBEG model, for $K/J=0$
and $r=0.3$. Same notation as in previous figure; $(c)$ stands for
$p=7$. Note the qualitative
change for $p>3$: a critical end point is present for $p=7$, as
well as reentrant behavior, with order-disorder-order transitions
as $\Delta/zJ$ is increased from small values.


\begin{thebibliography}{99}

\bibitem{barce} E. Vives, T. Cast\'an e P.-A. Lindg\aa rd,
Phys. Rev. B {\bf 53} (1996) 8915.
\bibitem{berker} See, for instance, K. Hui and A. N. Berker, Phys. Rev. Lett.
{\bf 62} (1989) 2507 or A. N. Berker, J. Appl. Phys. 
{\bf 70} (1991) 5941 and references therein.
\bibitem{branco} N. S. Branco and Beatriz Boechat, to be published in
Phys. Rev. B , November 1997.
\bibitem{cardy} J. Cardy and J. L. Jacobsen, preprint (cond-mat 9705038).
\bibitem{ajp} H. Falk, Am. J. Phys. {\bf 38} (1970) 858.
\bibitem{nsb} N. S. Branco and Luciano Bachmann, in preparation.

\end{thebibliography}
\end{document}